\documentclass[11pt,a4paper]{article}

\usepackage[utf8]{inputenc}
\usepackage[T1]{fontenc}
\usepackage[english]{babel}
\usepackage{geometry}
\usepackage{mathptmx}
\usepackage[expansion=false]{microtype}
\usepackage{amsmath}
\usepackage{booktabs}
\usepackage{tabularx}
\usepackage{fancyhdr}
\usepackage{titlesec}
\usepackage{enumitem}
\usepackage{xcolor}
\usepackage[hidelinks,breaklinks=true]{hyperref}
\hypersetup{
  pdftitle={From Digital Competence to Demonstrated Digital Capability: Positioning the International Digital Driving License Against DigComp and UNESCO Frameworks},
  pdfauthor={Ahmad Ghandour},
  pdfsubject={Digital competence, digital capability, behavioural assessment},
  pdfkeywords={digital competence, digital capability, digital literacy, DigComp, UNESCO, artificial intelligence competency, behavioural assessment, IDDL, human agency, digital citizenship}
}
\usepackage{url}
\makeatletter
\g@addto@macro\UrlBreaks{\UrlOrds}
\def\UrlBreaks{\do\/\do\-\do\.\do\_\do\a\do\b\do\c\do\d\do\e\do\f%
  \do\g\do\h\do\i\do\j\do\k\do\l\do\m\do\n\do\o\do\p\do\q\do\r%
  \do\s\do\t\do\u\do\v\do\w\do\x\do\y\do\z\do\0\do\1\do\2\do\3%
  \do\4\do\5\do\6\do\7\do\8\do\9}
\makeatother

\newcommand{\headerrule}{\rule{\textwidth}{0.4pt}}
\newcommand{\shorttitle}{From Digital Competence to Demonstrated Digital Capability}

\fancypagestyle{firstpage}{%
  \fancyhf{}
  
  \fancyhead[L]{\footnotesize\textsc{A Preprint}}
  \fancyhead[R]{\footnotesize\textsc{\today}}
  \fancyfoot[C]{\thepage}
}

\titleformat{\section}{\normalfont\large\bfseries}{\thesection}{0.6em}{}
\titleformat{\subsection}{\normalfont\normalsize\bfseries}{\thesubsection}{0.6em}{}
\titlespacing*{\section}{0pt}{1.6ex plus 0.6ex minus 0.2ex}{1.0ex plus 0.2ex}
\titlespacing*{\subsection}{0pt}{1.3ex plus 0.5ex minus 0.2ex}{0.8ex plus 0.2ex}

\newcommand{\bcite}[1]{\textbf{#1}}

\newenvironment{chain}
  {\begin{quote}\itshape\setlength{\parskip}{0.15em}}
  {\end{quote}}

\begin{document}

\thispagestyle{firstpage}

{\raggedright\LARGE\bfseries
From Digital Competence to Demonstrated Digital Capability:\\[0.25em]
Positioning the International Digital Driving License Against DigComp and UNESCO Frameworks\par}

\vspace{1.2em}

{\raggedright\large\bfseries Ahmad Ghandour\par}
\vspace{0.2em}
% TODO: insert institutional affiliation, postal address, email and ORCID before submission.
{\raggedright\normalsize Auckland, New Zealand\par}

\vspace{0.8em}
{\raggedright\normalsize\today\par}

\vspace{0.8em}
\headerrule

\vspace{1.2em}

\begin{center}
{\large\bfseries Abstract}
\end{center}
\vspace{-0.4em}
\begin{quote}
Digital competence frameworks define the knowledge, skills, attitudes and values required for participation in digital society. The European Digital Competence Framework for Citizens (DigComp) and UNESCO's digital literacy and artificial intelligence competency frameworks provide reference structures for policy, curriculum and competency development. The growth of generative and agentic AI raises a further question: does knowing what constitutes competent digital behaviour provide sufficient evidence that an individual can act competently in an authentic digital situation? This conceptual paper addresses that question through a comparative analysis of DigComp, UNESCO's Digital Literacy Global Framework, UNESCO's AI Competency Framework for Students, and the International Digital Driving License (IDDL). Rather than treating these frameworks as competing models, the paper distinguishes their functions. DigComp and UNESCO primarily define what individuals should know, understand, value and be able to do. IDDL addresses the subsequent evidentiary problem by assessing digital capability through behaviour observed during authentic digital tasks. Its Knowledge--Capability--Reflection model distinguishes knowing what should be done, demonstrating the required behaviour, and explaining the judgement behind it. IDDL integrates Digital Skills, Cybersecurity Awareness and AI Competency, and extends assessment to human agency under AI delegation. The paper distinguishes defined competence from demonstrated capability and positions IDDL as an interoperable behavioural assessment and certification layer that can operate downstream of international competency frameworks.
\end{quote}

\vspace{0.4em}
\begin{quote}
 UNESCO \textperiodcentered{} artificial intelligence competency \textperiodcentered{} behavioural assessment \textperiodcentered{} IDDL \textperiodcentered{} human agency \textperiodcentered{} digital citizenship
\end{quote}

\vspace{0.8em}
\headerrule
\vspace{0.6em}

%%%%%%%%%%%%%%%%%%%%%%%%%%%%%%%%%%%%%%%%%%%%%%%%%%%%%%%%%%%%%%%%%%%%%%%%%%%%%%

\section{Introduction}

Digital competence has moved from a specialist requirement to a condition of participation in contemporary society. Individuals search for information, communicate, transact, learn, work, and make decisions through digital systems. Artificial intelligence has intensified this dependence by shifting digital interaction from operating tools toward interacting with systems capable of generating information, recommending decisions, and increasingly undertaking actions on behalf of users.

International organisations have responded by developing frameworks that define the competencies required for digital participation. The European Commission's Digital Competence Framework for Citizens, commonly known as DigComp, has become one of the most influential reference frameworks. DigComp~2.2 organised digital competence around five competence areas and 21 competencies and supplemented them with more than 250 examples of knowledge, skills, and attitudes, including examples concerning artificial intelligence and emerging technologies (\bcite{Vuorikari et al., 2022}). DigComp~3.0 subsequently retained the basic architecture while updating competence statements, proficiency descriptions, learning outcomes, and the transversal treatment of AI competence (\bcite{Cosgrove \& Cachia, 2025}).

UNESCO has approached the problem from both digital literacy and AI competency perspectives. Its Digital Literacy Global Framework adapted and extended DigComp for international and cross-cultural use and supported the measurement of Sustainable Development Goal Indicator~4.4.2 (\bcite{Law et al., 2018}). More recently, UNESCO's AI Competency Framework for Students defined 12 competencies across four dimensions: human-centred mindset, ethics of AI, AI techniques and applications, and AI system design. These competencies progress through Understand, Apply, and Create levels (\bcite{Miao et al., 2024}).

These frameworks make substantial contributions to understanding what digital and AI competence entails. Yet the emergence of AI creates an assessment problem that deserves separate attention. An individual may understand cybersecurity principles but still respond unsafely to a sophisticated phishing attempt. A student may understand that AI outputs require verification but rely upon an inaccurate generated answer without checking it. A professional may recognise the importance of data privacy but disclose sensitive information when interacting with an AI system. More recently, an individual may understand that humans should retain authority over consequential decisions while nevertheless allowing an AI agent to take an external action without adequate supervision.

The distinction matters because knowledge about competent behaviour does not necessarily constitute evidence of competent behaviour.

This paper introduces the International Digital Driving License as a framework designed around that distinction. IDDL conceptualises digital capability through three interrelated forms of evidence: Knowledge, Capability, and Reflection. Rather than replacing international competency frameworks, IDDL seeks to operationalise selected competencies through authentic behavioural assessment.

The paper therefore addresses three questions.

\begin{enumerate}[leftmargin=2em,itemsep=0.25em,topsep=0.4em]
  \item How does IDDL conceptually differ from DigComp and UNESCO digital and AI competency frameworks?
  \item What additional evidence becomes available when digital competence is assessed through observable behaviour rather than knowledge or self-report alone?
  \item How might IDDL operate as an interoperable assessment and certification layer alongside established international competence frameworks?
\end{enumerate}

The central argument is that the next stage of digital competence development requires a clearer distinction between defined competence and demonstrated capability.

\section{Digital Competence as a Defined Construct}

DigComp provides a useful starting point because of its influence on digital competence policy, education, training, and assessment. DigComp~2.2 defines digital competence through five areas covering information and data literacy, communication and collaboration, digital content creation, safety, and problem solving. Across these areas sit 21 individual competencies supported by proficiency descriptors and examples of knowledge, skills, and attitudes (\bcite{Vuorikari et al., 2022}).

DigComp~2.2 also recognised the growing significance of AI. Its updated examples included citizens' interactions with AI systems and addressed issues such as data, algorithms, privacy, reliability, and human interaction with automated systems. The framework therefore should not be characterised as technologically outdated. Its contribution lies in providing a technology-aware but relatively stable competence architecture.

The publication of DigComp~3.0 strengthens this position. The revised framework maintains the overall structure while updating competence wording, revising proficiency levels, introducing learning outcomes, and integrating AI competence across the framework rather than treating AI as an isolated technical topic (\bcite{Cosgrove \& Cachia, 2025}).

This evolution illustrates an important feature of competence frameworks. Their primary function is normative and descriptive. They identify what constitutes competence and provide structures through which educators, governments, employers, and assessment bodies can interpret proficiency.

This function differs from demonstrating whether an individual performs competently when confronted with a particular digital situation.

\section{UNESCO and the Global Digital Literacy Perspective}

UNESCO's Digital Literacy Global Framework addressed the need for a reference structure capable of operating across diverse economic, cultural, and technological contexts. The framework drew substantially on DigComp while extending its scope for global measurement. It supported work surrounding SDG Indicator~4.4.2 and recognised that digital literacy must accommodate differences in technologies, occupations, infrastructure, and national contexts (\bcite{Law et al., 2018}).

The framework also demonstrates the difficulty of translating broad competency constructs into internationally comparable measures. Digital literacy encompasses activities ranging from basic device operation to information management, communication, content creation, safety, and problem solving. Measurement therefore requires decisions about what constitutes evidence and how competence should be inferred from that evidence.

This distinction between a reference framework and its measurement mechanism becomes important for IDDL. A framework can define a construct without prescribing a single assessment method. The existence of a competence statement does not determine whether it should be assessed through self-report, knowledge questions, performance tasks, portfolios, observation, or behavioural traces.

IDDL concentrates specifically on this measurement problem.

\section{UNESCO's AI Competency Framework}

The UNESCO AI Competency Framework for Students represents a significant development because it addresses competencies that extend beyond technical AI skills. It identifies four dimensions comprising a human-centred mindset, ethics of AI, AI techniques and applications, and AI system design. Twelve competency blocks span three progression levels of Understand, Apply, and Create (\bcite{Miao et al., 2024}).

The framework explicitly places human agency within AI competency. It emphasises critical judgement, accountability, responsible use, citizenship, and the need to maintain a human-centred relationship with AI. This orientation becomes increasingly important as AI systems move from generating content toward recommending and undertaking actions.

UNESCO's approach therefore shares substantial conceptual ground with IDDL. Both reject an interpretation of AI competency based solely on technical operation. Both recognise that competent AI use requires judgement, responsibility, verification, and human control.

Their primary difference concerns operationalisation.

UNESCO provides an international reference framework intended to guide curricula, learning outcomes, teaching, and assessment criteria. IDDL asks how these principles can become observable behavioural evidence during an assessment event.

Consider human agency. A learner may correctly explain that humans should remain responsible for consequential AI-assisted decisions. Such a response provides evidence of knowledge. A behavioural assessment introduces a different test. An AI agent can recommend an action, present incomplete evidence, or request authority to communicate externally. The assessment can then observe whether the individual verifies the evidence, establishes constraints, reserves decision authority, intervenes, refuses the action, or accepts the recommendation with justification.

The competency construct remains related. The evidentiary mechanism changes.

\section{The International Digital Driving License}

IDDL was developed around a simple proposition. Completion of learning and successful performance on a knowledge test do not by themselves establish that an individual can act competently within an authentic digital environment.

The framework therefore organises digital capability into three domains: Digital Skills, Cybersecurity Awareness, and AI Competency. These domains comprise 12 subdomains and 36 underlying concepts.

Its distinctive assessment architecture is the Knowledge--Capability--Reflection (K--C--R) model.

\begin{itemize}[leftmargin=2em,itemsep=0.25em,topsep=0.4em]
  \item \textbf{Knowledge} represents what the individual understands.
  \item \textbf{Capability} represents what the individual actually does.
  \item \textbf{Reflection} captures why the individual selected, rejected, modified, verified, or interrupted a course of action.
\end{itemize}

These components produce different forms of evidence. A participant might know that information generated by AI requires verification. During a task, however, the participant may accept an unsupported AI output. The knowledge evidence would therefore be positive while the capability evidence would contradict it. Reflection can provide further evidence about whether the behaviour resulted from deliberate judgement, misunderstanding, misplaced trust, or another reason.

IDDL consequently treats competence as an empirical claim that requires convergent evidence. This approach becomes important when certification makes a claim about what a person can do rather than merely what a person knows.

\section{From Knowledge Assessment to Behavioural Evidence}

Traditional assessment approaches frequently infer competence from responses to questions. Such methods remain useful and can measure conceptual understanding efficiently. Self-assessment provides another mechanism and can capture perceived confidence, experience, or attitudes. Neither approach necessarily demonstrates performance.

Behavioural assessment changes the unit of evidence.

Suppose a participant receives a message requesting urgent access to an institutional account. A knowledge question might ask which cybersecurity response is appropriate. The participant can select verification from several options.

An authentic task removes this abstraction. The message appears as part of the digital environment. The participant must decide what to inspect, whether to trust the sender, whether to follow a link, what evidence to gather, and whether to proceed.

The assessment therefore records behaviour rather than merely a statement about intended behaviour. This distinction can be represented as:

\begin{chain}
Defined competence $\rightarrow$ expected behaviour\\
Authentic task $\rightarrow$ observed behaviour\\
Observed behaviour + knowledge + reflection $\rightarrow$ capability inference\\
Repeated convergent evidence $\rightarrow$ certification decision
\end{chain}

The resulting certification claim is therefore narrower but potentially stronger. IDDL does not claim that a person possesses every possible digital competence. It seeks evidence that the person has demonstrated specified capability under defined assessment conditions.

\section{Comparative Framework Analysis}

The frameworks can be compared across purpose, scope, conceptualisation, assessment orientation, and certification logic.

DigComp functions primarily as a competence reference architecture. It provides a shared vocabulary and structured description of digital competence. UNESCO's Digital Literacy Global Framework extends this reference perspective toward global applicability and measurement. UNESCO's AI Competency Framework establishes an educational and normative architecture for responsible, human-centred AI competency. IDDL operates principally as a behavioural assessment and certification architecture.

Table~\ref{tab:comparison} summarises these functions as stated above.

\begin{table}[htbp]
\centering
\small
\caption{Primary function of each framework as characterised in this analysis.}
\label{tab:comparison}
\renewcommand{\arraystretch}{1.25}
\begin{tabularx}{\textwidth}{@{}lX@{}}
\toprule
\textbf{Framework} & \textbf{Primary function} \\
\midrule
DigComp & Competence reference architecture providing a shared vocabulary and structured description of digital competence. \\
UNESCO Digital Literacy Global Framework & Extension of the reference perspective toward global applicability and measurement. \\
UNESCO AI Competency Framework for Students & Educational and normative architecture for responsible, human-centred AI competency. \\
IDDL & Behavioural assessment and certification architecture. \\
\bottomrule
\end{tabularx}
\end{table}

The relationship can therefore be represented as:

\begin{chain}
International competence frameworks\\
$\downarrow$\\
Competency definitions and learning outcomes\\
$\downarrow$\\
Curriculum and learning experiences\\
$\downarrow$\\
Behavioural assessment\\
$\downarrow$\\
Evidence of demonstrated capability\\
$\downarrow$\\
Certification
\end{chain}

Under this interpretation, DigComp and UNESCO occupy upstream reference and educational layers. IDDL primarily occupies the assessment and certification layers. The frameworks are consequently potentially interoperable rather than mutually exclusive.

\section{Digital Skills, Cybersecurity, and AI as Integrated Capability}

A further characteristic of IDDL is the separation of Digital Skills, Cybersecurity Awareness, and AI Competency into three connected domains.

DigComp incorporates cybersecurity primarily within its broader Safety area, although cybersecurity also intersects other competence areas. UNESCO's AI framework addresses safety and responsible AI use within a broader ethical and human-centred architecture.

IDDL gives cybersecurity a dedicated domain because unsafe digital behaviour can invalidate otherwise competent task performance. A participant who completes a digital task successfully by exposing credentials or sensitive information has not demonstrated competent digital capability.

AI creates a similar issue. Effective prompt use alone does not establish AI competency. The user must also evaluate output, recognise limitations, protect sensitive information, make appropriate decisions, and determine when AI should or should not influence action.

Digital capability therefore emerges from interactions among operational ability, security judgement, and AI judgement rather than from isolated technical skills.

\section{Human Agency Under AI Delegation}

The development of agentic AI introduces a further challenge to existing competence models.

Generative AI largely required individuals to interact with systems through prompts and evaluate generated outputs. Agentic systems extend this relationship by planning activities, accessing tools, making recommendations, and undertaking authorised actions.

The assessment question therefore changes. The question is no longer only whether a person can use AI. It becomes whether a person can retain appropriate human agency while delegating tasks to AI.

Observable indicators can include verification before reliance, constraint setting, intervention, refusal, justified acceptance, protection of sensitive information, and reservation of consequential authority.

This distinction separates AI literacy from human agency under delegation. AI literacy concerns understanding, using, evaluating, and recognising the limitations of AI. Human agency under delegation concerns retaining judgement and responsibility when AI systems act with increasing operational autonomy.

UNESCO's emphasis on human agency and accountability provides a strong conceptual foundation for this development (\bcite{Miao et al., 2024}). IDDL provides one mechanism through which those principles can become assessable behaviours.

\section{The Driving Licence Analogy}

The conceptual logic of IDDL can be illustrated through driver licensing.

Road rules define what drivers need to know. Driver education develops knowledge and practical skills. A written examination can establish knowledge of those rules. None of these forms of evidence alone demonstrates that a person can drive safely in traffic. A practical driving examination therefore observes behaviour.

Digital environments increasingly share characteristics that make this distinction relevant. They are shared environments. Individual actions can affect other people and organisations. Poor decisions can produce financial, security, informational, and social consequences.

The IDDL proposition is therefore not that digital participation should replicate road regulation. The analogy concerns evidence. If certification claims that an individual can navigate digital environments competently, knowledge should form part of the evidence but should not constitute the whole of the evidence.

\section{Interoperability Rather Than Framework Competition}

Positioning IDDL as an alternative to DigComp or UNESCO would create an unnecessary framework competition. A more productive approach involves mapping IDDL behavioural constructs against established international competency definitions.

For example, IDDL Information Search behaviours can map to DigComp information and data literacy competencies. Cybersecurity behaviours can map to DigComp Safety competencies. AI evaluation, responsible use, and decision behaviours can map to relevant UNESCO AI competency dimensions. Human-agency behaviours can operationalise UNESCO principles concerning human-centred AI interaction and accountability.

Such mappings would enable institutions to retain established competence architectures while adopting behavioural methods for selected assessment purposes.

This approach also allows national adaptation. A country or institution could define digital literacy requirements through DigComp, UNESCO, or its own national framework and use IDDL-compatible behavioural tasks to assess whether selected competencies have been demonstrated.

IDDL would therefore function as an assessment protocol rather than another competing list of digital competencies.

\section{Discussion}

The comparison reveals a conceptual gap between defining competence and evidencing capability.

Existing international frameworks have made substantial progress in defining digital competence. DigComp provides a mature and evolving competence architecture. UNESCO has extended digital literacy toward global measurement and established a human-centred international framework for AI competency.

The challenge now concerns evidence.

The proliferation of generative AI makes knowledge-based assessment increasingly problematic because AI itself can produce plausible answers to knowledge questions. Agentic AI creates a deeper problem because digital competence increasingly involves decisions about delegation, supervision, verification, and intervention. Under these conditions, behavioural evidence becomes increasingly valuable.

This does not make knowledge assessment obsolete. Knowledge remains necessary. Nor does behavioural performance alone explain competence. A successful action may result from chance, imitation, or misunderstanding.

The K--C--R architecture addresses this problem by triangulating evidence. Knowledge establishes whether the participant understands the relevant principle. Capability establishes whether the participant applies it. Reflection establishes whether the participant can account for the judgement underlying the action. The convergence or divergence among these three forms of evidence provides a richer basis for capability inference.

\section{Implications for Research}

The framework creates several empirical questions.

Research should test whether K--C--R evidence predicts digital performance more effectively than knowledge tests or self-assessment alone. Studies should examine the relationship between stated knowledge and observed behaviour and identify conditions under which the two diverge.

Construct validity also requires examination. Researchers need to determine whether behavioural events provide stable evidence of underlying capability or merely reflect scenario-specific performance.

Cross-cultural validation will become important if IDDL operates internationally. Behaviour considered appropriate within one institutional or cultural context may not transfer directly to another. Mapping against DigComp and UNESCO provides a common reference architecture while allowing contextual variation in task design.

Human agency under AI delegation represents another research frontier. As agentic systems gain operational autonomy, researchers need validated measures of verification, constraint setting, intervention, refusal, authority reservation, and justified reliance.

These questions move digital literacy research beyond asking what people know about technology toward examining how people behave with technology.

\section{Conclusion}

DigComp, UNESCO, and IDDL address different parts of the digital competence problem.

DigComp provides an internationally influential architecture for defining digital competence. UNESCO extends digital literacy into global measurement and provides a human-centred framework for AI competency. IDDL addresses the evidentiary question that follows from these frameworks: can the individual demonstrate the capability that the framework says matters?

The distinction becomes increasingly important as AI mediates information, judgement, and action. Knowing that AI outputs should be verified differs from verifying them. Understanding privacy differs from protecting sensitive information during an AI interaction. Recognising human accountability differs from retaining authority when an AI agent proposes or undertakes consequential action.

Digital competence frameworks define what competent participation should entail. Behavioural assessment provides evidence about whether that competence can be demonstrated.

The proposed contribution of IDDL therefore lies not in replacing established frameworks but in connecting competency definitions with observable performance. Its Knowledge--Capability--Reflection architecture treats knowledge, behaviour, and reasoning as complementary sources of evidence and provides a foundation for capability-based certification.

The transition from digital competence to demonstrated digital capability may therefore represent an important next stage in digital citizenship education and assessment.

%%%%%%%%%%%%%%%%%%%%%%%%%%%%%%%%%%%%%%%%%%%%%%%%%%%%%%%%%%%%%%%%%%%%%%%%%%%%%%

\section*{References}
\addcontentsline{toc}{section}{References}

\begin{list}{}{%
  \setlength{\leftmargin}{1.5em}
  \setlength{\itemindent}{-1.5em}
  \setlength{\itemsep}{0.55em}
  \setlength{\parsep}{0pt}}

\item \textbf{Cosgrove, J., \& Cachia, R. (2025).} \emph{DigComp 3.0: European Digital Competence Framework --- Fifth edition} (EUR JRC144121). Publications Office of the European Union. \url{https://doi.org/10.2760/0001149}

\item \textbf{Law, N., Woo, D., de la Torre, J., \& Wong, G. (2018).} \emph{A global framework of reference on digital literacy skills for Indicator 4.4.2} (Information Paper No.~51). UNESCO Institute for Statistics. \url{https://uis.unesco.org/sites/default/files/documents/ip51-global-framework-reference-digital-literacy-skills-2018-en.pdf} (accessed 16 September 2026)

\item \textbf{Miao, F., Shiohira, K., \& Lao, N. (2024).} \emph{AI competency framework for students}. UNESCO. \url{https://doi.org/10.54675/JKJB9835}

\item \textbf{Vuorikari, R., Kluzer, S., \& Punie, Y. (2022).} \emph{DigComp 2.2: The Digital Competence Framework for Citizens --- With new examples of knowledge, skills and attitudes} (EUR 31006 EN, JRC128415). Publications Office of the European Union. \url{https://doi.org/10.2760/115376}

\end{list}

\end{document}